\documentclass[doublecol]{epl2}
\pdfoutput=1

\usepackage{subcaption}
\usepackage{amsmath}

\newcommand{\explainVerticalLine}{In the case of $S_{0}$ and $S_{0.5}$, the highlighted regions indicate radius values for which the sphere does not touch the crystal.}

\newcommand{\MainPlotScale}{0.3}

\title{Modelling Defect Cavities Formed in Inverse Three-Dimensional Rod-Connected Diamond Photonic Crystals}
\shorttitle{Modelling Defect Cavities Formed in Inverse 3D RCD Photonic Crystals} 

\author{M. P. C. Taverne\inst{1}\thanks{E-mail: \email{Mike.Taverne@bristol.ac.uk}}, Y.-L. D. Ho\inst{1}\thanks{E-mail: \email{Daniel.Ho@bristol.ac.uk}}, X. Zheng\inst{1}, S. Liu\inst{1}, L.-F. Chen\inst{1}, M. Lopez-Garcia\inst{1}, and J. G. Rarity\inst{1}\thanks{E-mail: \email{John.Rarity@bristol.ac.uk}}}
\shortauthor{M. P. C. Taverne \etal}

\institute{
    \inst{1} Department of Electrical and Electronic Engineering, University of Bristol - Merchant Venturers Building, Woodland Road, Bristol BS8 1UB, United Kingdom
}

\abstract{Defect cavities in 3D photonic crystal can trap and store light in the smallest volumes allowable in dielectric materials, enhancing non-linearities and cavity QED effects.
Here, we study inverse rod-connected diamond (RCD) crystals containing point defect cavities using plane-wave expansion and finite-difference time domain methods.
By optimizing the dimensions of the crystal, wide photonic band gaps are obtained.
Mid-bandgap resonances can then be engineered by introducing point defects in the crystal.
We investigate a variety of single spherical defects at different locations in the unit cell focusing on high-refractive-index contrast (3.3:1) inverse RCD structures;
quality factors ($Q$-factors) and mode volumes of the resonant cavity modes are calculated.
By choosing a symmetric arrangement, consisting of a single sphere defect located at the center of a tetrahedral arrangement, mode volumes \textless{}~0.06 cubic wavelengths are obtained, a record for high index cavities.}

\begin{document}

\maketitle

\section{Introduction}

An atomic crystal is formed from a periodic and systematic arrangement of atoms within which electrons feel a periodic potential.
This leads to wave vector specific electronic bandstructures and is the origin of electronic bandgaps in semiconductor materials.
Similarly, in a photonic crystal, the dielectric function varies periodically, leading to photonic band structures.
Thus, light can be manipulated by such a structure, when its periodicity is comparable to the wavelength of interest.
In the same way that localized electron states can be created within atomic crystal lattices by introducing defects, photons can be confined within defects in dielectric crystal lattices\cite{Okano2002, Tang2007, Ho2011:IEEE_JQE, Fu2013, Taverne2015}.
Studying these defect modes provides useful information for the development of versatile optoelectronic devices \cite{StevenJohnsonBook}.

Since 1987, when the concept of three-dimensional (3D) photonic crystal (PhCs) was considered for the modification of spontaneous emission by atoms \cite{Yablonovitch1987} and for strong localization of photons \cite{John1987}, research on PhCs became one of the most intensely studied subjects.
There have been successful attempts at fabricating 3D PhC structures using layer by layer lithographic techniques \cite{Aoki2003, Qi2004, Aoki2009, Ishizaki2009:NatureLetters, Ishizaki2013:NaturePhotonics}.
More recently, various high-refractive-index-contrast 3D PhC structures with complete PBGs have been fabricated using two-photon polymerization (2PP) based 3D lithography exploiting direct laser writing (DLW) with or without backfilling materials \cite{VonFreymann2010}.
Additionally, photonic diamond lattice structures such as rod-connected diamond (RCD) \cite{Datta1992, Chan1994, Maldovan2004}, are known to exhibit the largest full PBG \cite{Men2014:PhC-optimization}.
Such structures have been numerically investigated using the plane-wave expansion (PWE) method and finite-difference time domain (FDTD) calculations \cite{Imagawa2012}.
In our laboratory, we are actively pursuing 3D fabrication of such structures using DLW and have already seen partial bandgaps in 3D photonic crystals at $1.55 \mu m$ \cite{Chen2015}.

\section{Geometry and bandgaps: 3D RCD photonic crystal cavity design}

\begin{figure}[h!]
\onefigure[width=\columnwidth]{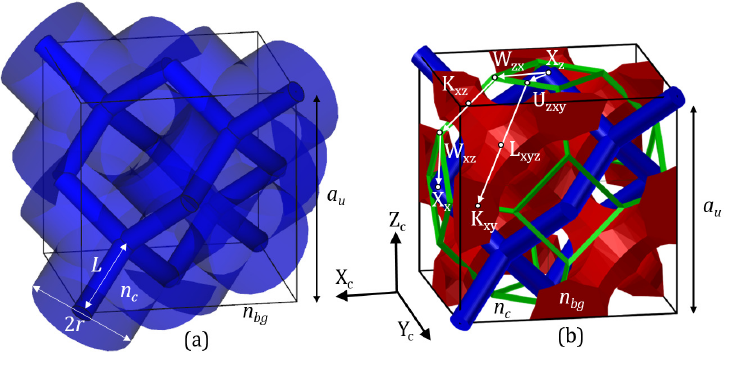}
\caption{(a) Cubic unit-cell of an RCD structure (transparent blue) with rods of radius $r$, and length $L =\sqrt{3} a_{u}/4$, where $a_{u}$ is the size of the cubic unit-cell.
(b) Cubic unit-cell of the inverse RCD structure (red) corresponding to the one shown in (a).
The large air rods of refractive index $n_{c}=1$ leave an RCD-like structure in the background material of refractive index $n_{bg}$.
The first Brillouin zone of the FCC lattice (green) and the k-points used in fig.~\ref{fig2_BandDiagram} are also shown.
In both figures, an RCD structure with smaller rod radius (blue) has been added for clarity.}
\label{fig1}
\end{figure}

\begin{figure}[h!]
\onefigure[width=0.75\columnwidth]{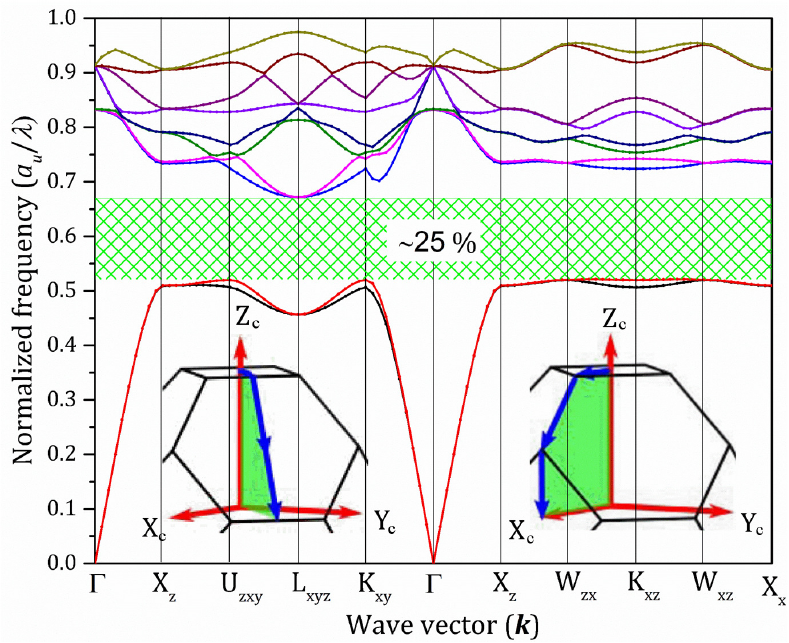}
\caption{Bandgap diagram of the optimal inverse RCD ($r=0.26a_{u}$).
The inset graphs show the movement described by the wavevector along the surface of the first brillouin zone in the corresponding left and right parts of the plot.}
\label{fig2_BandDiagram}
\end{figure}

In this paper, we are studying the inverse RCD structure as shown in fig.~\ref{fig1}.
The structure replaces the bonds in a diamond lattice with low-refractive-index rods surrounded by high-index material.
This can be visualized as a non-primitive cubic unit cell, containing 16 cylinders, which can be grouped into four regular tetrahedral arrangements.
Here, we investigate the potential to create high-$Q$, low mode volume cavities using the optical properties of single defects in 3D inverse RCD PhC cavities.
The photonic bandstructure of such RCD crystals was calculated using the MIT Photonic-Bands (MPB) package \cite{Johnson2001:mpb} for both non-inverse (high-index rods in air) and inverse (air rods in high-index backfill) structures.
Using a refractive index $n_{bg} = 3.3$ (simulating chalcogenide $Ge_{20}As_{20}Se_{14}Te_{46}$ or $GaP$) for the high-index material, a full photonic bandgap (PBG) is generally found between the second and third bands, as shown in fig.~\ref{fig2_BandDiagram}.

\begin{figure}[h!]
\onefigure[width=0.75\columnwidth]{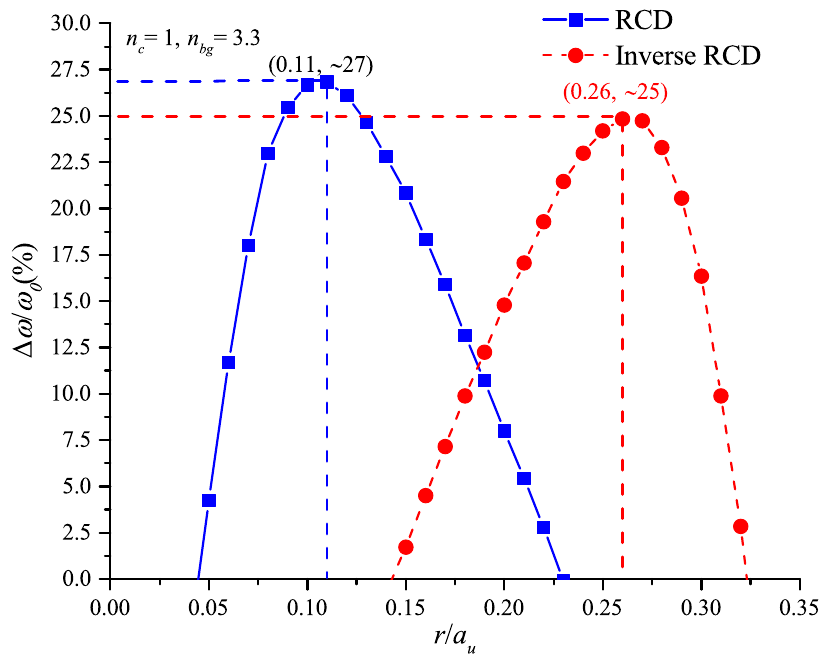}
\caption{Gap width to center frequency ratio $\Delta \omega / \omega_{0}$ (between band 2 and band 3) as a function of normalized rod radius $r / a_{u}$ for non-inverse and inverse RCD structures at refractive index contrast 3.3:1.}
\label{fig3_GapvsRadius}
\end{figure}

Figure~\ref{fig3_GapvsRadius} shows the gap-midgap ratio $\Delta \omega / \omega_{0}$ as a function of relative rod radius $r/a_{u}$, for the high-refractive-index contrast (3.3:1) inverse RCD structures.
The maximum gaps we find are $\Delta \omega / \omega_{0} \sim 27\%$ at $r_{RCD} = 0.11 a_{u}$ and $\Delta \omega / \omega_{0} \sim 25\%$ at $r_{inverse-RCD} = 0.26 a_{u}$ for non-inverse and inverse RCD respectively.
So while the non-inverse RCD offers a slightly wider bandgap, it is still possible to get a similar one in the inverse RCD, but by using a radius which is about 2.36 times larger.
This makes the inverse RCD structure easier to fabricate.

\begin{figure}[ht]
\onefigure[width=\columnwidth]{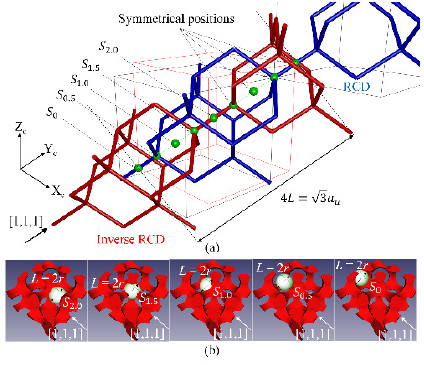}
\caption{(a) Simplified illustration of the RCD (blue) and inverse-RCD (red) both represented here using cylinders with small radii.
The green spheres along the [1,1,1] axis represent the positions of the studied defects.
(b) The spherical shape defects $S_{m}$ ($m$ = 0, 0.5, 1.0, 1.5, and 2.0), where $m$ corresponds to the location in fig.~\ref{fig4}(a).
The diameter of the defects is $D = 2 r_{d}$ (sphere radius varied from $r_{d} = 0.2 a_{u}$ to, $0.5 a_{u}$ in $0.025 a_{u}$ steps), $L = \sqrt{3} a_{u}/4$, is the length of the rods, and $a_{u}$ is the lattice constant of the RCD crystal as explained in fig.~\ref{fig1}.}
\label{fig4}
\end{figure}

\section{Calculating relevant parameters of the cavities}

In the following, we use FDTD simulations \cite{Railton:BristolFDTD} to calculate the relevant parameters for cavity structures placed at various points across the unit cell as illustrated in fig.~\ref{fig4}.
These are the cavity resonant wavelength $\lambda_{res}$, quality factor $Q$ and mode volume $V_{eff}$.
We assume $r / a_{u} = 0.26$ and refractive index contrast 3.3:1 as before.

The basis \{$X_{s}, Y_{s}, Z_{s}$\} used in the FDTD simulations is different from the conventional cubic unit-cell basis \{$X_{c}, Y_{c}, Z_{c}$\} used in figs.~\ref{fig1}, \ref{fig2_BandDiagram} and \ref{fig4}. The coordinates of $X_{s}, Y_{s}, Z_{s}$ in the conventional basis are: $X_{s} = [1, 0, -1] / \sqrt{2}$, $Y_{s} = [-1, 2, -1] / \sqrt{6}$, $Z_{s} = [1, 1, 1] / \sqrt{3}$.
This was done so that the $\Gamma-L$ direction ([1,1,1] axis in the conventional basis and direction of some of the cylinders) is aligned with the $Z_{s}$ direction.

The finite inverse RCD structures used for the FDTD simulations were created by truncating an infinite crystal using a cube of size $10a_{u} \times 10a_{u} \times 10a_{u}$ centred on the defect.
The refractive indices used were $n_{c}=1$ for the rods of the crystal, $n_{bg}=3.3$ for the background material and $n_{def}=3.3$ for the defect.
A non-homogeneous mesh of around $10^{7}$ cells was adapted to fit each simulation.
The simulations were conducted with a range of defect diameters $D = 2 r_{d}$ (varied from $r_{d} = 0.2 a_{u}$ to $0.5 a_{u}$ in $0.025 a_{u}$ steps), but at different positions within the RCD crystals.
The defects are labeled $S_{m}$ ($m$ = 0, 0.5, 1.0, 1.5, and 2.0) for a sphere, where $m$ corresponds to the location along the [1,1,1] axis of the crystal as illustrated in fig.~\ref{fig4}(a).
For each defect, three FDTD simulations (one for each direction: $X_{s}$, $Y_{s}$, $Z_{s}$) were run using a broadband dipole source placed in the defect.
After calculating the amplitude of the electric field over time for an inverse RCD with these defects, the $Q$-factors ($Q = \lambda_{res} / \Delta \lambda$) can then be estimated by analysing the resulting field decay in the frequency domain via the fast Fourier transform (FFT) and the filter diagonalisation method using the Harminv software \cite{Mandelshtam1997:harminv}.

\begin{figure*}
    \centering
    \begin{subfigure}{\MainPlotScale\linewidth}
        \centering
        \includegraphics[width=\linewidth]{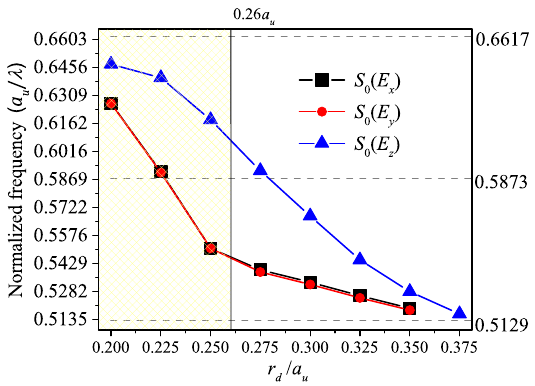}
        \caption{}
    \end{subfigure}
    \begin{subfigure}{\MainPlotScale\linewidth}
        \centering
        \includegraphics[width=\linewidth]{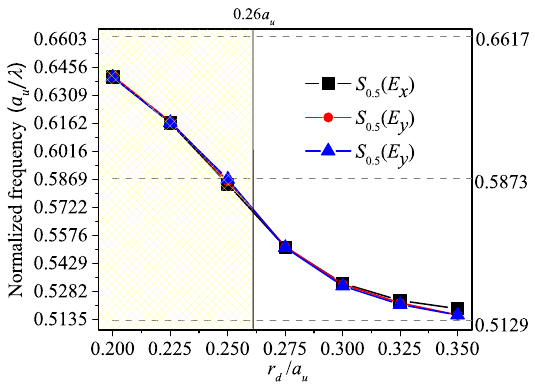}
        \caption{}
    \end{subfigure}
    \begin{subfigure}{\MainPlotScale\linewidth}
        \centering
        \includegraphics[width=\linewidth]{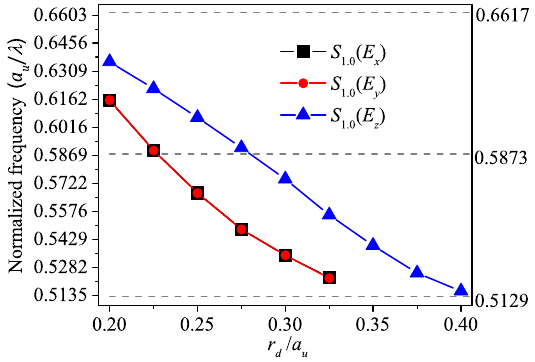}
        \caption{}
    \end{subfigure}
    
    \begin{subfigure}{\MainPlotScale\linewidth}
        \centering
        \includegraphics[width=\linewidth]{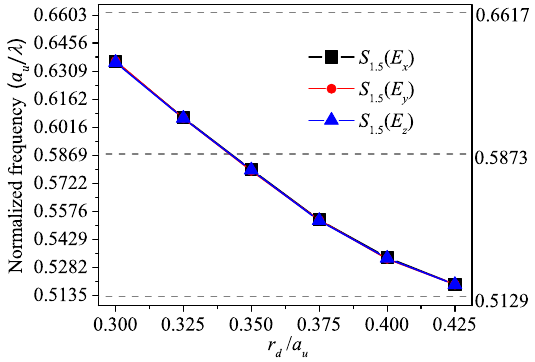}
        \caption{}
    \end{subfigure}
    \begin{subfigure}{\MainPlotScale\linewidth}
        \centering
        \includegraphics[width=\linewidth]{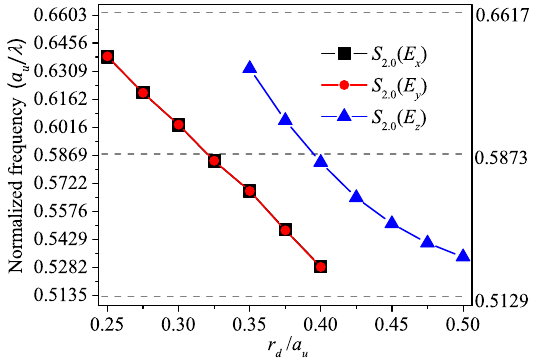}
        \caption{}
    \end{subfigure}
    \caption{The normalized frequency of resonance peaks ($ a_{u} / \lambda $) as a function of the normalized radius ($r_{d}/a_{u}$) of the defect for each defect position ($S_{m}$) with $E_{x}$-, $E_{y}$- and $E_{z}$-oriented dipole sources.
    The dashed lines indicate the limits of the full bandgap from $ a_{u} / \lambda \simeq 0.5129 $ to $0.6617$ and its midgap frequency $ a_{u} / \lambda \simeq 0.5873 $. \explainVerticalLine{}}
    \label{fig5}
\end{figure*}

\begin{figure*}
    \centering
    \begin{subfigure}{\MainPlotScale\linewidth}
        \centering
        \includegraphics[width=\linewidth]{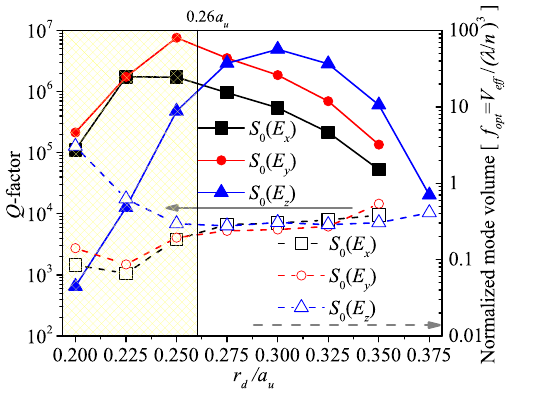}
        \caption{}
    \end{subfigure}
    \begin{subfigure}{\MainPlotScale\linewidth}
        \centering
        \includegraphics[width=\linewidth]{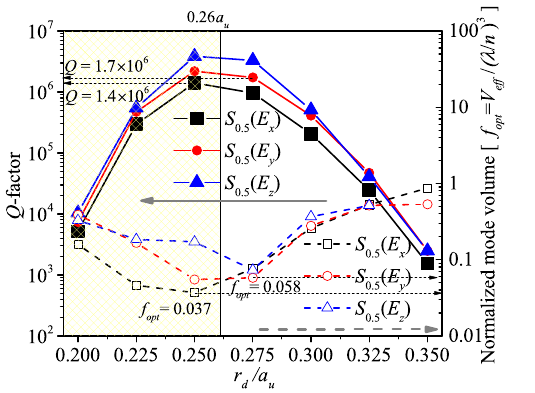}
        \caption{}
    \end{subfigure}
    \begin{subfigure}{\MainPlotScale\linewidth}
        \centering
        \includegraphics[width=\linewidth]{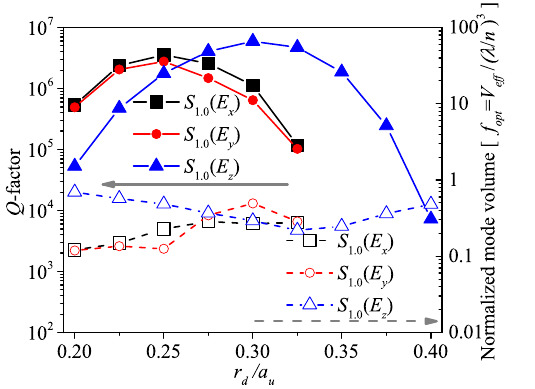}
        \caption{}
    \end{subfigure}
    
    \begin{subfigure}{\MainPlotScale\linewidth}
        \centering
        \includegraphics[width=\linewidth]{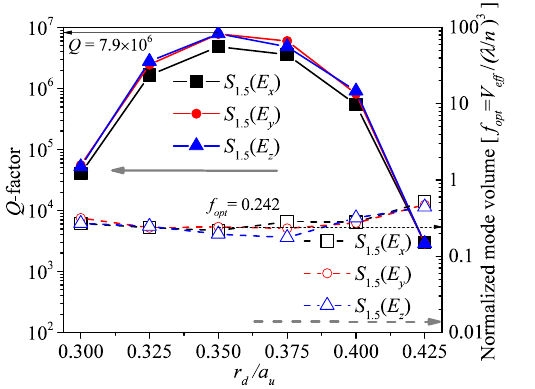}
        \caption{}
    \end{subfigure}
    \begin{subfigure}{\MainPlotScale\linewidth}
        \centering
        \includegraphics[width=\linewidth]{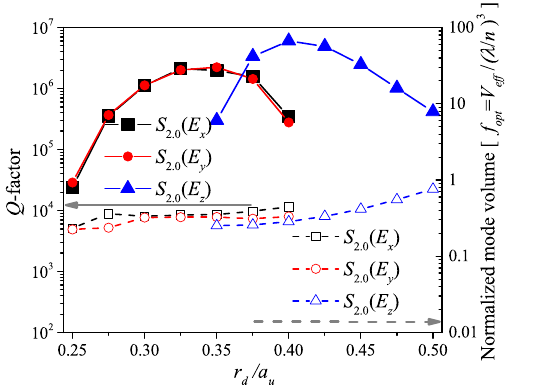}
        \caption{}
    \end{subfigure}
    \caption{$Q$-factors (solid lines) and the normalized mode volumes (dashed lines) obtained for the various normalized radii ($ r_{d} / a_{u} $) and locations ($S_{m}$ ($m$ = 0, 0.5, 1.0, 1.5, and 2.0), where $m$ corresponds to the location in fig.~\ref{fig4}) of sphere defects. \explainVerticalLine{}}
    \label{fig6_QVplots}
\end{figure*}

Figure~\ref{fig5} shows the normalized frequency of resonance peaks ($ a_{u} / \lambda $) as a function of defect radius for each defect position ($S_{m}$) within a full PBG between $ a_{u} / \lambda \sim 0.5129 $  and $0.6617$ (mid-gap frequency $ a_{u} / \lambda \simeq 0.5873 $).
For a variety of sphere defects ($S_{m}$), the corresponding $Q$-factors obtained for the different sphere radii are shown in fig.~\ref{fig6_QVplots}.
The defect radius ($ 0.35 a_{u} $) in the defect position $S_{1.5}$ gives a maximum $Q$-factor ($ Q \simeq 7.9 \times 10^{6} $) for the excitation in the $E_{y}$ and $E_{z}$-oriented dipole sources. While when the resonance peak ($ a_{u} / \lambda $) is too close to the lower band-edge (see fig.~\ref{fig5}), the sphere defect ($S_{0}$) gives a much lower $Q$-factor (down to $Q \sim 1000 $) as seen in fig.~\ref{fig6_QVplots}(a).

\begin{figure}
\begin{subfigure}{0.49\linewidth}
\centering
\includegraphics[width=\linewidth]{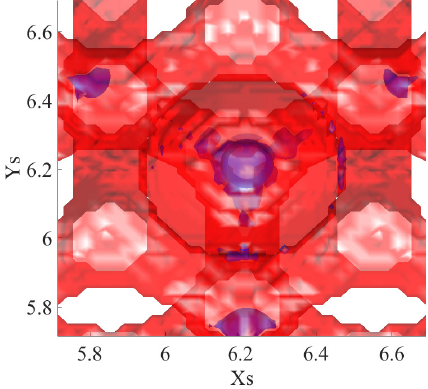}
\caption{$S_{0.5}$, $E_{z}$-dipole}
\label{fig7a}
\end{subfigure}
\begin{subfigure}{0.49\linewidth}
\centering
\includegraphics[width=\linewidth]{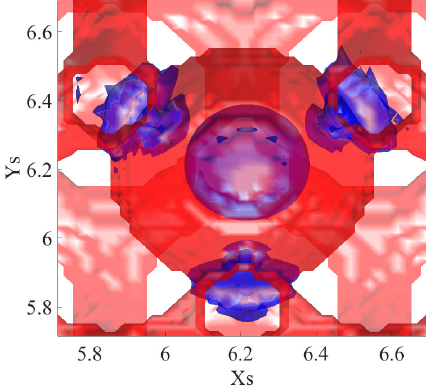}
\caption{$S_{1.5}$, $E_{z}$-dipole}
\label{fig7b}
\end{subfigure}

\begin{subfigure}{0.49\linewidth}
\centering
\includegraphics[width=\linewidth]{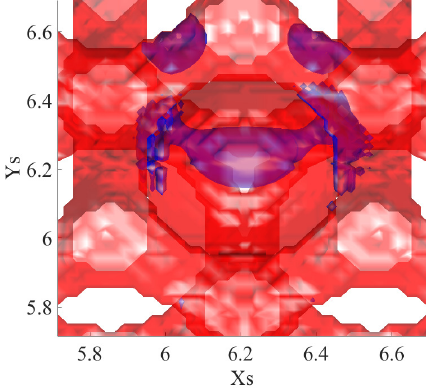}
\caption{$S_{0.5}$, $E_{x}$-dipole}
\label{fig7c}
\end{subfigure}
\begin{subfigure}{0.49\linewidth}
\centering
\includegraphics[width=\linewidth]{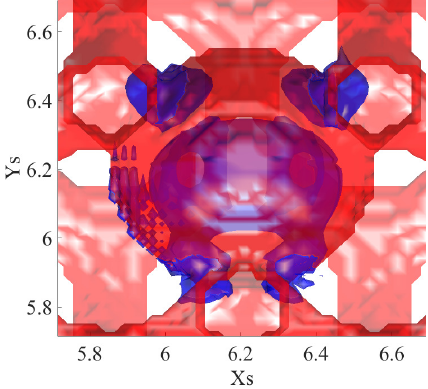}
\caption{$S_{1.5}$, $E_{x}$-dipole}
\label{fig7d}
\end{subfigure}

\begin{subfigure}{0.49\linewidth}
\centering
\includegraphics[width=\linewidth]{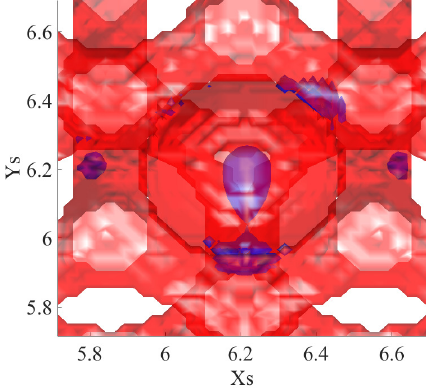}
\caption{$S_{0.5}$, $E_{y}$-dipole}
\label{fig7e}
\end{subfigure}
\begin{subfigure}{0.49\linewidth}
\centering
\includegraphics[width=\linewidth]{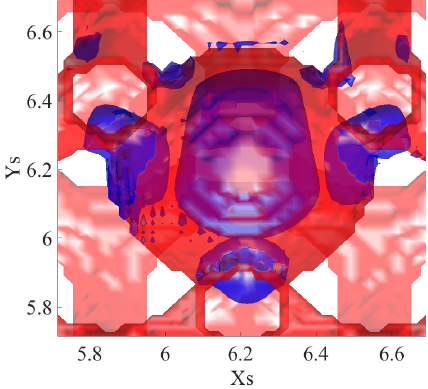}
\caption{$S_{1.5}$, $E_{y}$-dipole}
\label{fig7f}
\end{subfigure}
\caption{Isosurfaces of the material (red) and energy density (blue) distributions for the defects $S_{0.5}$ with radius $r_{d} = 0.275 a_{u}$ (a,c,e) and $S_{1.5}$ with radius $r_{d} = 0.35 a_{u}$ (b,d,f), after an initial broadband Gaussian modulated sinewave excitation pulse in the $Z_{s}=Z_{c}$ (i.e. $\Gamma-L$) (a,b), $X_{s}$ (c,d) and $Y_{s}$ (e,f) directions.
The dimensions are normalized by $a_{u}$.}
\label{fig7_isosurfaces}
\end{figure}

Having determined the resonant frequency, a cavity mode on resonance can be visualized using single frequency snapshots.
The confinement of the energy density distribution [$\varepsilon (|E|^{2})$] of the sphere defects ($S_{0.5}$,$r=0.275a_{u}$) and ($S_{1.5}$,$r=0.35a_{u}$) is illustrated in fig.~\ref{fig7_isosurfaces} for dipoles oriented along the $X_{s}$, $Y_{s}$ and $Z_{s}$ directions.
In all cases, the field is strongly localized to the defect with some outlier peaks in surrounding high index regions.
An effective mode volume ($V_{eff}$) of the cavity modes can be calculated from FDTD simulation results using \cite{Coccioli1998, Boroditsky1999}:
\begin{equation}
\label{eq.1}
V_{eff}=\frac{\iiint\varepsilon(r)|E(r)|^{2}dr^{3}}{\left[\varepsilon|E(r)|^{2}\right]_{max}}
\end{equation}

A useful figure of merit is the dimensionless effective volume $f_{opt}$, which is the effective cavity mode volume $V_{eff}$ normalized to the cubic wavelength of the resonant mode in a medium of refractive index $n$, defined as:
\begin{equation}
\label{eq.2}
f_{opt}=\frac{V_{eff}}{(\lambda/n)^{3}}
\end{equation}

The corresponding mode volumes are shown in fig.~\ref{fig6_QVplots}.
The smallest mode volume was $V_{eff} \sim 0.037 (\lambda_{res}/n)^{3}$.
It was obtained for the defect $S_{0.5}$ (a sphere radius $r = 0.25 a_{u}$ at the center of a "tetrahedral arrangement") excited by an $E_{x}$-oriented dipole source, with a resonance at $a_{u} / \lambda \simeq 0.5845$ of $Q$-factor $1.4 \times 10^{6}$.
However, for this radius, the defect sphere does not touch the crystal and hangs freely in space.
Although one could postulate using ultra-thin supporting "strings", this is not really feasible at our chosen optical frequency.
If only feasible defects are considered, the smallest mode volume is still obtained for the same defect, but with a slightly larger radius $r = 0.275 a_{u}$ and an $E_{y}$-oriented dipole source.
In this case, $V_{eff} \sim 0.058 (\lambda_{res}/n)^{3}$, $a_{u} / \lambda \simeq 0.55$ and $Q \sim 1.7 \times 10^{6}$.

The obtained $f_{opt}$ values can now be used to estimate the expected coupling strength $g_{R}$ of the cavity mode with a quantum emitter placed inside the defect by using the following formula \cite{Andreani1999,Ho2011:IEEE_JQE,Taverne2015}:
\begin{equation}
\label{eq.3}
g_{R}=\left(\frac{3c_{0}\gamma_{os}n_{def}}{8\pi f_{opt}\lambda_{os}n_{os}}\right)^{1/2}
\end{equation}
where $c_{0}$ is the speed of light in vacuum, $\gamma_{os}$ and $\lambda_{os}$ are the spontaneous emission rate and wavelength of the considered quantum emitter and $n_{os}$ is the refractive index directly surrounding the emitter.
The weak and strong coupling criteria $ 4 g_{R} / ( \kappa_{uc} + \gamma_{os} ) $, where $ \kappa_{uc} = 2 \pi c_{0} / (\lambda_{os} Q) $ is the decay rate of the uncoupled cavity, can also be calculated.
Here, we consider coupling to the zero-phonon line (ZPL) of diamond $NV^{-}$ centers, for which $n_{os} = 2.4$, $ \lambda_{os} = 637 nm $ and $ \gamma_{os} = \gamma_{ZPL} \sim 4\% \cdot \gamma_{total} = 3.3 \times 10^{6} \: rad/s $ because emission into the ZPL constitutes only $4\%$ of the total spontaneous emission \cite{Ho2011:IEEE_JQE,Taverne2015}.

\begin{figure*}
    \centering
    \begin{subfigure}{\MainPlotScale\linewidth}
        \centering
        \includegraphics[width=\linewidth]{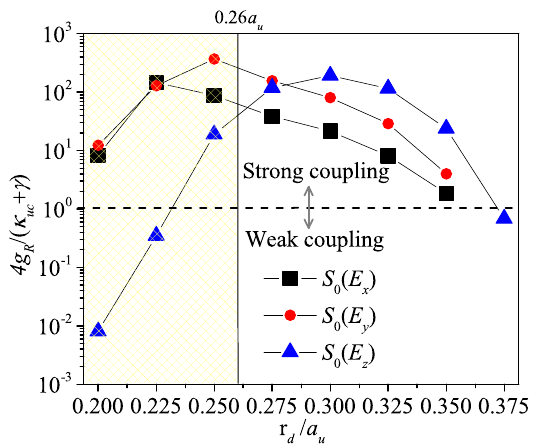}
        \caption{}
    \end{subfigure}
    \begin{subfigure}{\MainPlotScale\linewidth}
        \centering
        \includegraphics[width=\linewidth]{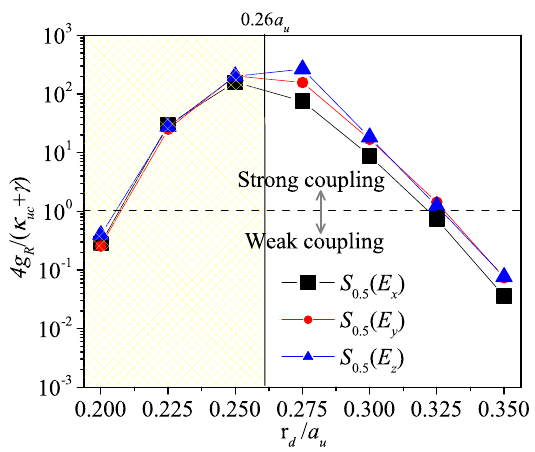}
        \caption{}
    \end{subfigure}
    \begin{subfigure}{\MainPlotScale\linewidth}
        \centering
        \includegraphics[width=\linewidth]{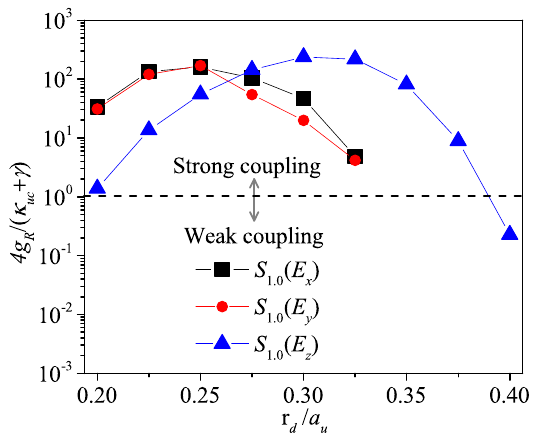}
        \caption{}
    \end{subfigure}
    
    \begin{subfigure}{\MainPlotScale\linewidth}
        \centering
        \includegraphics[width=\linewidth]{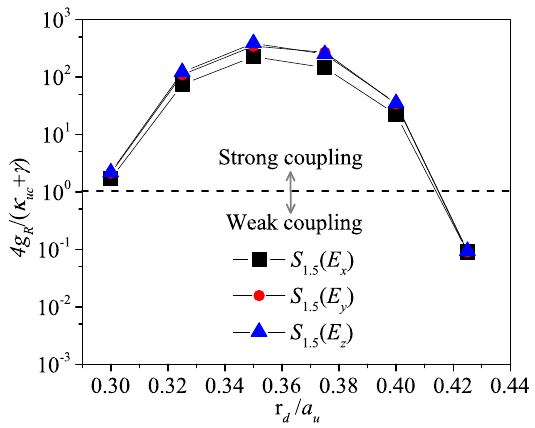}
        \caption{}
    \end{subfigure}
    \begin{subfigure}{\MainPlotScale\linewidth}
        \centering
        \includegraphics[width=\linewidth]{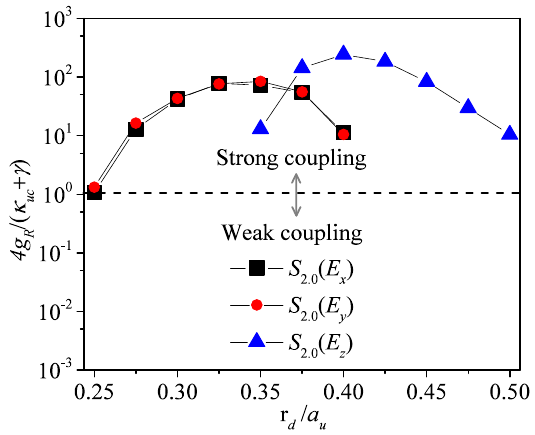}
        \caption{}
    \end{subfigure}
    \caption{Strong/weak coupling criteria $ 4 g_{R} / (\kappa_{uc} + \gamma) $ of the studied inverse RCD defects $S_{m}$ ($m$ = 0, 0.5, 1.0, 1.5, and 2.0) against the normalized radius ($ r_{d} / a_{u} $) of sphere defects for coupling to a diamond $NV^{-}$ center. \explainVerticalLine{}}
    \label{fig8}
\end{figure*}

Figure~\ref{fig8} shows the strong/weak coupling criteria $ 4 g_{R} / ( \kappa_{uc} + \gamma_{os} ) $ of the studied inverse RCD defect positions $S_{m}$ ($m$ = 0, 0.5, 1.0, 1.5, and 2.0) in the case of coupling to a diamond $NV^{-}$ center.
Almost all considered defects have $  4 g_{R} / ( \kappa_{uc} + \gamma_{os} ) > 1 $ and should therefore lead to strong coupling, i.e. a splitting of the cavity line, even at low excitation powers resulting in a double peaked emission spectrum.
In fact the strong coupling criterion is over 100 times its threshold value for specific radius values in all defect positions.

It is noticeable that the curves for the different excitation directions ($E_{x}$, $E_{y}$ and $E_{z}$) are similar in the cases of $S_{0.5}$ and $S_{1.5}$, when looking at figs.~\ref{fig5}, \ref{fig6_QVplots}, \ref{fig8}.
However, for $S_{0.0}$, $S_{1.0}$ and $S_{2.0}$, the $E_{z}$ excitations lead to different values than the $E_{x}$ and $E_{y}$ excitations.
Moreover, the $S_{0.5}$ and $S_{1.5}$ positions are at the center of tetrahedral arrangements and therefore have a higher symmetry than the other positions, which have more of an axial symmetry along the [1,1,1] direction, i.e. the $Z_{s}$ axis used in the FDTD simulations.
This makes sense if we look at fig.~\ref{fig4}.
Hence, in the former case, it would seem that all excitation directions are able to excite the same (or equivalent via rotation) modes, while in the latter cases, the $E_{z}$ excitation is able to excite a different mode.
However, the results of Q-factors, mode volumes and coupling strength in figs. \ref{fig6_QVplots} and \ref{fig8} are partly not in agreement with the above facts (eg., the Q-factors for $S_{0}$ ($E_{x}$) and $S_{0}$ ($E_{y}$) are considerably different in fig.~\ref{fig6_QVplots}(a)).
Fig.~\ref{fig7_isosurfaces} also shows that, in the case of $S_{0.5}$ and $r_{d} = 0.275 a_{u}$, each of the dipole orientations leads to different field distributions, despite all resonant modes having almost the same frequency $a_{u} / \lambda \simeq 0.55$.
The excited modes are therefore sometimes distinct, even if the resonance frequencies are close.
We ascribe this as due to the different symmetries seen in the planes orthogonal to the x, y and z dipoles (i.e. the $X_{s}$, $Y_{s}$ and $Z_{s}$ planes).

The $S_{1.5}$ and $S_{2.0}$ positions require larger radiuses than the other positions to support resonant modes (larger than $0.300 a_{u}$ and $0.250 a_{u}$ respectively).
This is due to the fact that in those cases, the high-index sphere defects are positioned inside high-index regions of the photonic crystal.
It is then necessary to use a radius large enough to actually modify the local material distribution.
$S_{1.5}$ being at the intersection of four "inverse rods" (see fig.~\ref{fig4}) requires a larger radius than $S_{2.0}$, which is at the center of a single "inverse rod", with the "inverse rods" being smaller at their centres than at their ends, as can be seen in fig.~\ref{fig1}(b).

\section{Conclusion}

In this paper, 3D inverse RCD PhCs formed in high-index-contrast materials (GaP or chalcogenide) were investigated and a maximum PBG of $ \sim 25 \% $ at $ (r / a_{u})_{opt} = 0.26 $ was reported by using the plane wave expansion method.
Moreover, a variety of sphere defect positions, located along the [1,1,1] axis going through rods of the 3D RCD structures, were investigated and the $Q$-factors and mode volumes of their cavity modes calculated using the FDTD method.
Among the considered feasible defects, the sphere defect ($S_{0.5}$) gives the best results for an excitation in the $E_{y}$ direction with a mode volume $ V_{eff} \sim 0.058 (\lambda_{res} / n_{GaP})^{3} $, while still having high $Q$-factors $ Q = 1.7 \times 10^{6} $.
Thus, this is smaller than the dielectric cavity mode volumes obtained for non-inverse RCD ($ \sim 0.09 (\lambda_{res} / n_{Si})^{3} $ where $ n_{Si} = 3.6 $ \cite{Imagawa2012}) and GaP based woodpile PhC cavities ($ \sim 0.1 (\lambda_{res} / n_{GaP})^{3} $, where $ n_{GaP} = 3.3 $ \cite{Taverne2015}).
Furthermore, it is about one order of magnitude smaller than the mode volume obtained for the 2D PhC high index defect microcavity ($ \sim 0.25 (\lambda_{res} / n_{InGaAs})^{3} $ where $ n_{InGaAs} = 3.46 $ \cite{Coccioli1998, Boroditsky1999}).
This is the smallest high index cavity volume that has been seen\revision{, although smaller normalised cavity volumes have been reported in low index cavities in 1D crystals (slotted nanobeam)\cite{Robinson2005,Ryckman2012:APL}}.
Such a high-$Q$ cavity with small mode volume could allow the investigation of strong coupling cavity QED effects on non-ideal quantum emitters such as diamond $NV^{-}$ color centers \cite{Vahala2003,Reiserer2015,Vos2015:book-chapter}.

Compared to woodpiles\cite{Fu2013}, the required inverse RCD crystal size to reach Q-factors of one million is $\sim 33\%$ smaller ($\sim 1500 a_{u}^3$ versus $10^3 a_{u}^3$).
Additionally, the plane orthogonal to $\Gamma-L$ exhibits a hexagonal symmetry similar to the one encountered in 2D hexagonal crystal slabs.
This similarity can be exploited to transfer existing waveguide designs from such geometries into RCD and inverse RCD strucures.

The main drawback compared to woodpiles is the fabrication difficulty.
However, this can be overcome by using DLW\cite{Chen2015} and other innovative fabrication techniques\cite{Aoki2009}.

\acknowledgments
This work was carried out using the computational facilities of the Advanced Computing Research Centre, University of Bristol, Bristol, U.K. and we acknowledge financial support from the ERC advanced grant 247462 QUOWSS and EPSRC grant EP/M009033/1.

\bibliographystyle{eplbib}
\bibliography{main}

\begin{thebibliography}{10}
\expandafter\ifx\csname url\endcsname\relax\def\url#1{\texttt{#1}}\fi

\bibitem{Okano2002}
\Name{Okano M., Chutinan A. \and Noda S.} \REVIEW{Phys. Rev.
  B}{66}{2002}{165211}.
\newline\url{https://doi.org/10.1103/PhysRevB.66.165211}

\bibitem{Tang2007}
\Name{Tang L. \and Yoshie T.} \REVIEW{Optics Express}{15}{2007}{17254}.
\newline\url{https://doi.org/10.1364/OE.15.017254}

\bibitem{Ho2011:IEEE_JQE}
\Name{Ho Y.~D., Ivanov P.~S., Engin E., Nicol M. F.~J., Taverne M. P.~C.,
  {Chengyong Hu}, Cryan M.~J., Craddock I.~J., Railton C.~J. \and Rarity J.~G.}
  \REVIEW{IEEE Journal of Quantum Electronics}{47}{2011}{1480}.
\newline\url{https://doi.org/10.1109/JQE.2011.2170404}

\bibitem{Fu2013}
\Name{Fu J., Tandaechanurat A., Iwamoto S. \and Arakawa Y.} \REVIEW{physica
  status solidi (c)}{10}{2013}{1457}.
\newline\url{http://dx.doi.org/10.1002/pssc.201300282}

\bibitem{Taverne2015}
\Name{Taverne M. P.~C., Ho Y.-l.~D. \and Rarity J.~G.} \REVIEW{Journal of the
  Optical Society of America B}{32}{2015}{639}.
\newline\url{https://doi.org/10.1364/JOSAB.32.000639}

\bibitem{StevenJohnsonBook}
\Name{Joannopoulos J.~D., Johnson S.~G., Winn J.~N. \and Meade R.~D.}
  \Book{{Photonic Crystals: Molding the Flow of Light}} 2nd Edition (Princeton
  University Press) 2008.
\newline\url{http://ab-initio.mit.edu/book/}

\bibitem{Yablonovitch1987}
\Name{Yablonovitch E.} \REVIEW{Physical Review Letters}{58}{1987}{2059}.
\newline\url{https://doi.org/10.1103/PhysRevLett.58.2059}

\bibitem{John1987}
\Name{John S.} \REVIEW{Physical Review Letters}{58}{1987}{2486}.
\newline\url{https://doi.org/10.1103/PhysRevLett.58.2486}

\bibitem{Aoki2003}
\Name{Aoki K., Miyazaki H.~T., Hirayama H., Inoshita K., Baba T., Sakoda K.,
  Shinya N. \and Aoyagi Y.} \REVIEW{Nature Materials}{2}{2003}{117}.
\newline\url{https://doi.org/10.1038/nmat802}

\bibitem{Qi2004}
\Name{Qi M., Lidorikis E., Rakich P.~T., Johnson S.~G., Joannopoulos J.~D.,
  Ippen E.~P. \and Smith H.~I.} \REVIEW{Nature}{429}{2004}{538}.
\newline\url{https://doi.org/10.1038/nature02575}

\bibitem{Aoki2009}
\Name{Aoki K.} \REVIEW{Applied Physics Letters}{95}{2009}{191910}.
\newline\url{https://doi.org/10.1063/1.3264088}

\bibitem{Ishizaki2009:NatureLetters}
\Name{Ishizaki K. \and Noda S.} \REVIEW{Nature}{460}{2009}{367}.
\newline\url{http://dx.doi.org/10.1038/nature08190}

\bibitem{Ishizaki2013:NaturePhotonics}
\Name{Ishizaki K., Koumura M., Suzuki K., Gondaira K. \and Noda S.} \REVIEW{Nat
  Photon}{7}{2013}{133}.
\newline\url{http://dx.doi.org/10.1038/nphoton.2012.341}

\bibitem{VonFreymann2010}
\Name{von Freymann G., Ledermann A., Thiel M., Staude I., Essig S., Busch K.
  \and Wegener M.} \REVIEW{Advanced Functional Materials}{20}{2010}{1038}.
\newline\url{https://doi.org/10.1002/adfm.200901838}

\bibitem{Datta1992}
\Name{Datta S., Chan C.~T., Ho K.~M. \and Soukoulis C.~M.} \REVIEW{Phys. Rev.
  B}{46}{1992}{10650}.
\newline\url{https://doi.org/10.1103/PhysRevB.46.10650}

\bibitem{Chan1994}
\Name{Chan C., Datta S., Ho K. \and Soukoulis C.} \REVIEW{Physical Review
  B}{50}{1994}{1988}.
\newline\url{https://doi.org/10.1103/PhysRevB.50.1988}

\bibitem{Maldovan2004}
\Name{Maldovan M. \and Thomas E.~L.} \REVIEW{Nature Materials}{3}{2004}{593}.
\newline\url{https://doi.org/10.1038/nmat1201}

\bibitem{Men2014:PhC-optimization}
\Name{Men H., Lee K. Y.~K., Freund R.~M., Peraire J. \and Johnson S.~G.}
  \REVIEW{Optics Express}{22}{2014}{22632}.
\newline\url{https://doi.org/10.1364/OE.22.022632}

\bibitem{Imagawa2012}
\Name{Imagawa S., Edagawa K. \and Notomi M.} \REVIEW{Applied Physics
  Letters}{100}{2012}{151103}.
\newline\url{https://doi.org/10.1063/1.4704182}

\bibitem{Chen2015}
\Name{Chen L., Taverne M. P.~C., Zheng X., Lin J.-D., Oulton R., Lopez-Garcia
  M., Ho Y.-L.~D. \and Rarity J.~G.} \REVIEW{Optics Express}{23}{2015}{26565}.
\newline\url{https://doi.org/10.1364/OE.23.026565}

\bibitem{Johnson2001:mpb}
\Name{Johnson S. \and Joannopoulos J.} \REVIEW{Optics Express}{8}{2001}{173}.
\newline\url{https://doi.org/10.1364/OE.8.000173}

\bibitem{Railton:BristolFDTD}
\Name{Railton C. \and Hilton G.} \REVIEW{IEEE Transactions on Antennas and
  Propagation}{47}{1999}{707}.
\newline\url{https://doi.org/10.1109/8.768811}

\bibitem{Mandelshtam1997:harminv}
\Name{Mandelshtam V.~A. \and Taylor H.~S.} \REVIEW{The Journal of Chemical
  Physics}{107}{1997}{6756}.
\newline\url{https://doi.org/10.1063/1.475324}

\bibitem{Coccioli1998}
\Name{Coccioli R., Boroditsky M., Kim K., Rahmat-Samii Y. \and Yablonovitch E.}
  \Book{{Smallest possible electromagnetic mode volume in a dielectric cavity}}
  in proc. of \Book{IEE Proceedings Optoelectronics} Vol. 145 (IET) 1998 p.
  391.
\newline\url{http://dx.doi.org/10.1049/ip-opt:19982468}

\bibitem{Boroditsky1999}
\Name{Boroditsky M., Vrijen R., Krauss T., Coccioli R., Bhat R. \and
  Yablonovitch E.} \REVIEW{Journal of Lightwave Technology}{17}{1999}{2096}.
\newline\url{https://doi.org/10.1109/50.803000}

\bibitem{Andreani1999}
\Name{Andreani L., Panzarini G. \and G{\'{e}}rard J.-M.} \REVIEW{Physical
  Review B}{60}{1999}{13276}.
\newline\url{https://doi.org/10.1103/PhysRevB.60.13276}

\bibitem{Robinson2005}
\Name{Robinson J.~T., Manolatou C., Chen L. \and Lipson M.} \REVIEW{Physical
  Review Letters}{95}{2005}{143901}.
\newline\url{https://doi.org/10.1103/PhysRevLett.95.143901}

\bibitem{Ryckman2012:APL}
\Name{Ryckman J.~D. \and Weiss S.~M.} \REVIEW{Applied Physics
  Letters}{101}{2012}{071104}.
\newline\url{http://dx.doi.org/10.1063/1.4742749}

\bibitem{Vahala2003}
\Name{Vahala K.~J.} \REVIEW{Nature}{424}{2003}{839}.
\newline\url{https://doi.org/10.1038/nature01939}

\bibitem{Reiserer2015}
\Name{Reiserer A. \and Rempe G.} \REVIEW{Rev. Mod. Phys.}{87}{2015}{1379}.
\newline\url{https://doi.org/10.1103/RevModPhys.87.1379}

\bibitem{Vos2015:book-chapter}
\Name{Vos W.~L. \and Woldering L.~A.} \Book{{Cavity quantum electrodynamics
  with three-dimensional photonic bandgap crystals}} in \Book{Light
  Localisation and Lasing: Random and Quasi-Random Photonic Structures}, edited
  by \Name{Ghulinyan M. \and Pavesi L.} (Cambridge University Press, Cambridge)
  2015 pp. 180--213.
\newline\url{http://doc.utwente.nl/99061/}

\end{thebibliography}

\end{document}